\documentclass[journal,onecolumn]{IEEEtran}

\usepackage{mathpazo}
\usepackage{times}
\usepackage{amsmath}
\usepackage{amsfonts}
\usepackage{latexsym}
\usepackage{amssymb}
\usepackage{cite}

\usepackage{upref}
\usepackage{theorem}
\usepackage{graphicx}
\usepackage{psfrag}
\usepackage{verbatim}
\usepackage{algorithm}
\usepackage{algorithmic}

\usepackage[usenames]{color}

\theoremstyle{plain}
\theorembodyfont{\normalfont\slshape}

\newtheorem{thm}{Theorem$\!$}
\newenvironment{theorem}
{\begin{thm}\hspace*{-1ex}{\bf.}}{\end{thm}}

\newtheorem{lem}[thm]{Lemma$\!$}
\newenvironment{lemma}{\begin{lem}\hspace*{-1ex}{\bf.}}{\end{lem}}

\newtheorem{prop}[thm]{Proposition$\!$}

\newtheorem{cor}[thm]{Corollary$\!$}
\newenvironment{corollary}{\begin{cor}\hspace*{-1ex}{\bf.}}{\end{cor}}

\newtheorem{defn}[thm]{Definition$\!$}
\newenvironment{definition}{\begin{defn}\hspace*{-1ex}{\bf.}}{\end{defn}}

\newtheorem{rem}[thm]{Remark$\!$}

\newtheorem{xmpl}[thm]{Example$\!$}
\newenvironment{example}{\begin{xmpl}\hspace*{-1ex}{\bf.}}{\hfill$\Box$\end{xmpl}}

\newtheorem{cnstr}{Construction$\!$}

\newenvironment{construction}{\begin{cnstr}\hspace*{-1ex}{\bf.}}{\hfill$\Box$\end{cnstr}}

\newcounter{enumrom}
\renewcommand{\theenumrom}{(\roman{enumrom})}

\makeatletter
\renewcommand{\@endtheorem}{\endtrivlist}
\makeatother

\makeatletter
\renewcommand{\thefigure}{{\@arabic\c@figure}}
\renewcommand{\fnum@figure}{{\bf Figure\,\thefigure}}
\makeatother

\newcommand{\cB}{\mathcal{B}}

\newcommand{\cL}{\mathcal{L}}

\newcommand{\mathset}[1]{\left\{#1\right\}}
\newcommand{\abs}[1]{\left|#1\right|}
\newcommand{\ceilenv}[1]{\left\lceil #1 \right\rceil}
\newcommand{\floorenv}[1]{\left\lfloor #1 \right\rfloor}
\newcommand{\parenv}[1]{\left( #1 \right)}
\newcommand{\sparenv}[1]{\left[ #1 \right]}

\newcommand{\be}[1]{\begin{equation}\label{#1}}
\newcommand{\ee}{\end{equation}}

\renewcommand{\leq}{\leqslant}

\renewcommand{\geq}{\geqslant}

\renewcommand{\Bbb}{\mathbb}

\newcommand{\Cref}[1]{Co\-ro\-lla\-ry\,\ref{#1}}

\renewcommand{\Bbb}{\mathbb}

\newcommand{\N}{{\Bbb N}}

\newcommand{\Z}{{\Bbb Z}}

\DeclareMathOperator{\id}{Id}

\newcommand{\dmin}{d_{\mathrm{min}}}
\newcommand{\rmin}{r_{\mathrm{min}}}
\newcommand{\eqdef}{\triangleq}
\newcommand{\rt}{\tilde{r}}
\newcommand{\Bt}{\tilde{B}}
\newcommand{\Tt}{\tilde{T}}
\newcommand{\spn}[1]{\left\langle #1 \right\rangle}
\newcommand{\ccyc}{C^{\mathrm{cyc}}}
\newcommand{\cid}{C^{\id}}
\newcommand{\Mcyc}{M^{\mathrm{cyc}}}

\newcommand{\Mid}{M^{\id}}

\newcommand{\lmin}{\cL_{\min}}
\newcommand{\lmax}{\cL_{\max}}

\outer\def\proclaim #1. #2\par{\medbreak
 \noindent{\bf#1.\enspace}{\sl#2\par}%
 \ifdim\lastskip<\medskipamount \removelastskip\penalty55\medskip\fi}

\makeatletter
\newcommand*{\bmodp}{%
  \nonscript\mskip-\medmuskip\mkern5mu%
  \mathbin{\operator@font mod}^+ \penalty900\mkern5mu%
  \nonscript\mskip-\medmuskip
}
\makeatother

\begin{document}


\title{\textbf{Infinity-Norm Permutation Covering Codes\\ from Cyclic
    Groups}}

\author{\large
Ronen Karni and
Moshe~Schwartz,~\IEEEmembership{Senior Member,~IEEE}%
\thanks{Ronen Karni is with the Department
   of Electrical and Computer Engineering, Ben-Gurion University of the Negev,
   Beer Sheva 8410501, Israel
   (e-mail: karniron@post.bgu.ac.il).}%
\thanks{Moshe Schwartz is with the Department
   of Electrical and Computer Engineering, Ben-Gurion University of the Negev,
   Beer Sheva 8410501, Israel
   (e-mail: schwartz@ee.bgu.ac.il).}%
\thanks{This work was supported in part by the Israel Science Foundation (ISF) grant No.~130/14.}
}

\maketitle

\begin{abstract}
  We study covering codes of permutations with the
  $\ell_\infty$-metric. We provide a general code construction, which
  uses smaller building-block codes. We study cyclic transitive groups
  as building blocks, determining their exact covering radius, and
  showing linear-time algorithms for finding a covering codeword. We
  also bound the covering radius of relabeled cyclic transitive groups
  under conjugation.
\end{abstract}



\section{Introduction}

\IEEEPARstart{C}{oding} over permutations appears in the literature as
early as the works \cite{Sle65,ChaKur69}. In a typical setting, the
symmetric group of permutations, $S_n$, is endowed with a distance
function, $d:S_n\times S_n\to \N_0$, to create a metric. An
\emph{error correcting code} is then defined as a set $C\subseteq
S_n$, the elements of which are called \emph{codewords}, such that
$d(f,g)\geq \dmin$, for all $f,g\in C$, $f\neq g$. The largest such
$\dmin$ is called the minimum distance of the code. It is also well
known that $C$ induces a packing of the space, $S_n$, by disjoint
balls of radius $\floorenv{(\dmin-1)/2}$, the \emph{packing radius},
centered at the codewords.

In this work, we are interested in the dual problem of
covering. Instead of packing balls, we are interested in the smallest
radius of balls, centered at the codewords, such that their union
covers the entire space. This radius is called the \emph{covering
  radius} of the code. Equivalently, we are looking for the smallest
$\rmin\in\N_0$ such that every $f\in S_n$ has a codeword $g\in C$ with
$d(f,g)\leq \rmin$.

Covering codes over permutations have only recently been studied in
depth, starting with the work of \cite{CamWan05}, and following with
\cite{Qui06,KeeKu06}, all of which only use the Hamming distance over
permutations. In \cite{CamWan05}, the exact size of covering codes
over $S_n$ and covering radius $n-1$ is found, and bounds are given on
the size of covering codes with smaller covering radius.  In
\cite{KeeKu06}, the authors present a randomized construction for a
code and use a certain frequency parameter to bound the covering
radius of the code. A survey of error-correcting codes and covering
codes over permutations is given in \cite{Qui06}.

Motivated by applications to information storage in non-volatile
memories, the rank-modulation scheme was recently suggested
\cite{JiaMatSchBru09}, in which information is stored in the form of
permutations. The relevant permutation metrics for this scheme are
mainly the $\ell_\infty$-metric and Kendall's-$\tau$ metric. Thus, we
have works studying error-correcting codes
\cite{JiaSchBru10,TamSch10,KloLinTsaTze10,BarMaz10,TamSch12,FarSkaMil13,MazBarZem13,ZhoSchJiaBru15,YehSch16a},
Gray codes and snake-in-the-box codes
\cite{YehSch12b,HorEtz14,ZhaGe16,Hol16,WanFu16}, and related
combinatorial questions \cite{Klo09,SchTam11,Klo11}.

Covering codes over permutations with the $\ell_\infty$-metric have
only been studied in \cite{WanMazWor15,FarSchBru16a}. In
\cite{WanMazWor15}, various connections between different metrics over
permutations were found, thus enabling code construction in the
$\ell_\infty$-metric based on codes in other metrics. Additionally,
bounds on code parameters were given, which were later improved in
\cite{FarSchBru16a}, together with an explicit direct code
construction.

The main contribution of this paper is a generalization of the code
construction from \cite{FarSchBru16a}. This generalization requires
smaller building-block covering codes. We study one such
building-block code in detail -- a cyclic transitive group of
$S_n$. We derive the exact covering radius of this group, as well as
bound its covering radius after relabeling (conjugation). We also
provide linear-time covering-codeword algorithm for the codes.

The paper is organized as follows. In Section \ref{sec:prelim} we
introduce formal definitions and notations used throughout the
paper. Section \ref{sec:cover} is devoted to the derivation of the
covering radius of the naturally labeled cyclic transitive group. In
Section \ref{sec:code} we describe the generalized code construction,
as well as linear-time algorithms associated with it. We then turn in
Section \ref{sec:relabel} to studying relabeling of the building-block
code and finding bounds on its covering radius. We conclude in Section
\ref{sec:disc} by discussing the results and suggesting open problems.


\section{Notations and Definitions}
\label{sec:prelim}

For $m,m'\in\N$, we denote $[m,m']\eqdef\mathset{m,m+1,\dots,m'}$, as
well as $[m]\eqdef[1,m]$. For ease of notation, we write $m \bmodp n$
to denote the unique $r\in [n]$ such that $n$ divides $m-r$. We then
define the \emph{cyclic interval}
\[ [m,m']\bmodp n \eqdef \mathset{m\bmodp n, (m+1)\bmodp n, \dots,m' \bmodp n}.\]

The symmetric group of permutations is denoted by $S_n$. As will be
evident later, it is important for us to fix the permuted
elements. Thus, a permutation $f\in S_n$ is a bijection between $[n]$
and itself. We shall use either a one-line notation for permutations,
where $f=[f_1,f_2,\dots,f_n]$ denotes a permutation mapping $i\mapsto
f_i$ for all $i\in[n]$, or a cycle notation $f=(f_1,f_2,\dots,f_k)$
where $f$ maps $f_i\mapsto f_{(i+1)\bmodp k}$ for all $i\in[k]$. If
$f,g\in S_n$ are two permutations, their composition is denoted by
$fg$, where $(fg)(i)=f(g(i))$ for all $i\in[n]$. The identity
permutation is denoted by $\id$.

The metric of interest in this work is the $\ell_\infty$-metric,
sometimes also called the Chebyshev metric. The distance function in
this metric, denoted $d_\infty: S_n\times S_n\to \N_0$, is defined for
all $f,g\in S_n$ by
\[ d_\infty(f,g) \eqdef \max_{i\in[n]}\abs{f(i)-g(i)}.\]
Since this will be the only distance function of interest, we shall
drop the $\infty$ subscript and use only $d$. We note that for all
$f,g\in S_n$, we have $d(f,g)\leq n-1$. It is well known (e.g., see
\cite{DezHua98}) that $d$ is right invariant (but not left invariant),
i.e., for all $f,g,h\in S_n$,
\[ d(fh,gh)=d(f,g).\]

A \emph{code} $C$ is simply a subset $C\subseteq S_n$. Sometimes $C$
will also be a subgroup of $S_n$, in which case we may refer to $C$
as a \emph{group code}. For such a code $C\subseteq S_n$, and $f\in S_n$,
we define the distance between $f$ and $C$ by
\[ d(f,C)\eqdef \min_{g\in C} d(f,g).\]
The main object of study in this work is now defined.

\begin{definition}
  An $(n,M,r)$ \emph{covering code} is a subset $C\subseteq S_n$, such
  that $\abs{C}=M$ and $d(f,C)\leq r$ for all $f\in S_n$, and $r$ is
  the minimal integer with this property.
\end{definition}

Given an $(n,M,r)$ covering code $C$, we call $r(C)\eqdef r$ the
\emph{covering radius} of $C$. In an asymptotic setting it will be
useful to define the \emph{rate} of the code, and its \emph{normalized covering radius} by
\begin{align*}
  R(C)&\eqdef\frac{\log_2 M}{n}, &\qquad\qquad \rho(C)&\eqdef \frac{r}{n-1}.
\end{align*}

The main focus throughout this paper involves cyclic groups. Since the
distance function crucially depends on the permuted elements, we need
to define a ``natural'' description of these group. Additionally, to
avoid degenerate cases, we shall only examine transitive cyclic
groups. We therefore give the following definition.

\begin{definition}
  \label{def:natgn}
  For all $n\in\N$, the (natural, transitive) cyclic group, denoted
  $G_n\leq S_n$, is the group generated by the permutation
  $(1,2,\dots,n)$, i.e.,
  \begin{equation}
    \label{eq:gn}
    G_n \eqdef \spn{(1,2,\dots,n)} = \mathset{ (1,2,\dots,n)^k : k\in\Z }.
  \end{equation}
\end{definition}

It will additionally be helpful to have a notation for permutations
that are close enough to the code. If $f,g\in S_n$ and $d(f,g)\leq
\rt$, we say $f$ is \emph{$\rt$-covered by $g$}, and otherwise, we say
$f$ is \emph{$\rt$-exposed by $g$}. If $C\subseteq S_n$ is a code, and
$f\in S_n$ is $\rt$-covered by at least one $g\in C$, i.e.,
$d(f,C)\leq\rt$, we say $f$ is \emph{$(\rt,C)$-covered}. Otherwise,
$f$ is $\rt$-exposed by every $g\in C$, and we say $f$ is
\emph{$(\rt,C)$-exposed}. In the latter case, for every $g\in C$,
there exists $i\in[n]$ such that $\abs{f(i)-g(i)}>\rt$, and we say
that the mapping $i\mapsto f(i)$ is \emph{$\rt$-exposed by $g$}.


\section{The Covering Radius of the Cyclic Group}
\label{sec:cover}

In this section we determine the covering radius of the natural
transitive cyclic group. This will later be used as a component in a
more general construction for covering codes. We first present two
bounds on the covering radius, that nearly agree. We then close the
small gap to obtain the exact covering radius.

Throughout this section, let $G_n\leq S_n$ denote the natural
transitive cyclic group of \eqref{eq:gn}. Since for $n=1,2$, we have
$G_n=S_n$, we trivially have $r(G_1)=r(G_2)=0$. Thus, in what follows
we focus on $n\geq 3$.

If $f\in S_n$ is some permutation, $H\leq S_n$ a subgroup, and
$\rt\in\N$, we define
\[ A^H_{i\mapsto f(i)}\eqdef \mathset{ h^{-1}(1) ~:~ \text{$i\mapsto f(i)$ is $\rt$-exposed by $h\in H$}}.\]
Since we will be mainly interested in the case of $H=G_n$, we define
\[A_{i\mapsto f(i)}\eqdef A^{G_n}_{i\mapsto f(i)}.\]
We also define the two sets
\begin{align*}
  B & \eqdef [n-\rt-1] & T&\eqdef [\rt+2,n],
\end{align*}
for the bottom and top parts of the range $[n]$. In these definitions,
to keep the notation simple, the dependence on $n$ and $\rt$ is
implicit. Some simple observations are formalized in the next two
lemmas.

\begin{lemma}
  \label{lem:entrycover}
  Let $f\in S_n$ be any permutation, and $\rt\in\N$. If $H\leq S_n$ is
  a transitive group, $\abs{H}=n$, then $f$ is $(\rt,H)$-exposed if
  and only if
  \begin{equation}
    \label{eq:triv1}
    \bigcup_{i\in[n]}A^H_{i\mapsto f(i)}=[n].
  \end{equation}
\end{lemma}
\begin{IEEEproof}
  If \eqref{eq:triv1} holds, since $\abs{H}=n$, it follows that every
  $h\in H$ $\rt$-exposes $f$, hence $f$ is $(\rt,H)$-exposed. In the
  other direction, if $f$ is $(\rt,G_n)$-exposed, then every $g\in
  G_n$ $\rt$-exposes some mapping $i\mapsto f(i)$. Since
  $\bigcup_{g\in G_n}\mathset{g^{-1}(1)}=[n]$, the claim follows.
\end{IEEEproof}

\begin{lemma}
  \label{lem:exposedset}
  Let $\rt,n\in\N$, $\rt\geq\frac{n}{2}-1$, and $H\leq S_n$ a
  transitive subgroup, $\abs{H}=n$. Then for all $i,j\in[n]$,
  \[\abs{A^H_{i\mapsto j}}=\begin{cases}
  n-\rt-j & j\in B=[n-\rt-1], \\
  j-\rt-1 & j\in T=[\rt+2,n],\\
  0 & \text{otherwise.}
  \end{cases}\]
  In particular, for $H=G_n$, for all $j_B\in B$, $j_T\in
  T$, and $i_B,i_T\in[n]$,
  \begin{align*}
    A_{i_B\to j_B} &= [i_B+1,i_B+n-\rt-j_B] \bmodp n, \\
    A_{i_T\to j_T} &= [i_T-j_T+\rt+2,i_T] \bmodp n.
  \end{align*}
\end{lemma}
\begin{IEEEproof}
  Consider the first claim. If $i\mapsto j$, $j\in B$, is
  $\rt$-exposed by some $h\in H$, then $h(i)\in [j+\rt+1,n]$. Thus,
  since $H$ is transitive and $\abs{H}=n$, there are exactly $n-\rt-j$
  such $h\in H$, proving the claim regarding the size of
  $A^H_{i\mapsto j}$.

  Additionally, when considering $H=G_n\eqdef\spn{(1,2,\dots,n)}$,
  we know $h^{-1}(1)=(i_B-h(i_B)+1)\bmodp n$. Combining this with the
  range of $h(i_B)$ we get
  \[A_{i_B\to j_B} = [i_B+1,i_B+n-\rt-j_B] \bmodp n.\]
  The rest of the claims, involving $T$, $i_T$, and $j_T$, are proven
  symmetrically.
\end{IEEEproof}

We can now prove an upper bound on the covering radius of $G_n$.
\begin{lemma}
  \label{lem:upperbound}
  For all $n\in\N$, $n\geq 3$,
  \[ r(G_n)\leq n-\ceilenv{\frac{\sqrt{4n+1}-1}{2}}.\]
\end{lemma}
\begin{IEEEproof}
  Let $f\in S_n$ be any permutation, and consider any $\rt\in\N$ in
  the range $\frac{n}{2}-1\leq\rt\leq n-1$. Using Lemma
  \ref{lem:exposedset},
  \begin{equation}
    \label{eq:union}
    \abs{\bigcup_{i\in n} A_{i\mapsto f(i)}}\leq 2\sum_{i=1}^{n-\rt-1}i=
    (n-\rt-1)(n-\rt).
  \end{equation}
  By Lemma \ref{lem:entrycover}, if
  \begin{equation}
    \label{eq:rttop}
    (n-\rt-1)(n-\rt) < n,
  \end{equation}
  then $f$ is $(\rt,G_n)$-covered. The smallest value of $\rt$ that
  satisfies \eqref{eq:rttop} is
  \[ \rt=n-\ceilenv{\frac{\sqrt{4n+1}-1}{2}},\]
  and since for any $\rt$ that satisfies \eqref{eq:rttop} we have
  $r(G_n)\leq\rt$, we obtain the desired bound.
\end{IEEEproof}

We now move on to a lower bound on the covering radius of $G_n$.

\begin{lemma}
  \label{lem:lowerbound}
  For all $n\in\N$, $n\geq 3$,
  \[r(G_n)\geq n-\floorenv{\frac{\sqrt{4n+1}+1}{2}}.\]
\end{lemma}
\begin{IEEEproof}
  By simple inspection, $r(G_3)=1$, agreeing with the claim. We
  therefore focus on the remaining case of $n\geq 4$. For convenience
  we define
  \begin{align*}
    a &\eqdef \floorenv{\frac{\sqrt{4n+1}+1}{2}} &
    \rt &\eqdef n-a-1.
  \end{align*}

  The proof strategy is the following: we shall define a permutation
  $f_0\in S_n$ and show that $f_0$ is $(\rt,G_n)$-exposed. It would
  then follow that $r(G_n)\geq \rt+1=n-a$, which would complete the
  proof.

  We construct a permutation $f_0\in S_n$ as follows:
  \begin{align}
    \label{eq:f0}
    f_0(i) &\eqdef
    \begin{cases}
      n-a+k &i=\binom{k+1}{2}, k\in [a], \\
      a-\ell+1 &i=2\binom{a+1}{2}-1-\binom{\ell+1}{2}, \ell\in [a], \\
      \text{arbitrary} &\text{otherwise},
    \end{cases}\\
    &=\begin{cases}
    j_T &i=\binom{a-(n-j_T)+1}{2}, j_T\in T, \\
    j_B &i=2\binom{a+1}{2}-1-\binom{a-(j_B-1)+1}{2}, j_B\in B, \\
    \text{arbitrary} &\text{otherwise},
    \end{cases} \nonumber
  \end{align} 
  for all $i\in[n]$, and where arbitrary entries are set in a way that
  completes $f_0$ to a permutation.

  We first contend that $f_0$ is well defined. We note that since
  $n\geq 4$ we have $B\cap T=\emptyset$, so the values in the range of
  $f_0$ are distinct. As for the domain, the first two cases of \eqref{eq:f0}
  are disjoint, since otherwise we would have $k,\ell\in[a]$ such that
  \[ \binom{k+1}{2}+\binom{\ell+1}{2}=2\binom{a+1}{2}-1.\]
  This obviously does not hold for $k=\ell=a$, as well as
  $k,\ell\in[a-1]$. The only remaining case is when
  $\mathset{k,\ell}=\mathset{a,a-1}$. However, it is easy to verify
  that
  \[ \binom{a+1}{2}+\binom{a}{2}=2\binom{a+1}{2}-1,\]
  only when $a=1$, which is never the case when $n\geq 4$. Hence,
  $f_0$ is indeed a well defined permutation.

  We now proceed with showing that $f_0$ is $(\rt,G_n)$-exposed. By
  examining the first case of \eqref{eq:f0} and using Lemma
  \ref{lem:exposedset}, we obtain for all $j_T\in T$,
  \begin{align*}
    \bigcup_{j_T\in T} A_{f_0^{-1}(j_T)\mapsto j_T} &= \bigcup_{k\in[a]}
    \sparenv{\binom{k+1}{2}-k+1,\binom{k+1}{2}}\bmodp n \\
    &= \sparenv{\binom{a+1}{2}}\bmodp n.
  \end{align*}
  Symmetrically, let $\ell'\in[a]$ be the smallest integer such that
  \[ 2\binom{a+1}{2}-1-\binom{\ell'+1}{2}\leq n.\]
  Then by Lemma \ref{lem:exposedset},
  \begin{align*}
    &\bigcup_{j_B\in B} A_{f_0^{-1}(j_B)\mapsto j_B} \\
    &\quad = \bigcup_{\ell\in[\ell',a]}
    \sparenv{2\binom{a+1}{2}-\binom{\ell+1}{2},2\binom{a+1}{2}-\binom{\ell+1}{2}+\ell-1}\bmodp n \\
    &\quad = \sparenv{\binom{a+1}{2},2\binom{a+1}{2}-1-\binom{\ell'}{2}}\bmodp n.
  \end{align*}
  We now note that
  \begin{align*}
    2\binom{a+1}{2}-1 &= \floorenv{\frac{\sqrt{4n+1}+1}{2}} \parenv{\floorenv{\frac{\sqrt{4n+1}+1}{2}}+1}-1\\
    &> \frac{\sqrt{4n+1}-1}{2}\cdot \frac{\sqrt{4n+1}+1}{2} -1 \\
    &= n-1,
  \end{align*}
  and since the expression on the left-hand side is an integer, we get
  \[ 2\binom{a+1}{2}-1 \geq n.\]
  Additionally, the choice of $\ell'$ ensures that also
  \[2\binom{a+1}{2}-1-\binom{\ell'}{2} \geq n.\]
  It then follows that
  \[ \bigcup_{i\in[n]} A_{i\mapsto f_0(i)} = [n],\]
  and by Lemma \ref{lem:entrycover}, $f_0$ is $(\rt,G_n)$-exposed.
\end{IEEEproof}
    
\begin{example}
For $n=7$, from \eqref{eq:f0} we get
\[ f_0 = [5,?,6,?,1,7,?], \]
where $?$ represents entries that can be mapped arbitrarily so as to
complete a permutation from $S_7$. Denote $g=(1,2,3,4,5,6,7)$, so that
$G_7=\spn{g}$. Table \ref{tab:G7f0Exp} shows the entries of $f_0$
which were mapped to $B\cup T$, and the permutations $g^k\in G_7$ by
which they are $3$-exposed. It also details the relevant $A_{i\mapsto f_0(i)}$ sets. We conclude that $r(G_7)\geq 4$, since $f_0$ is
$(3,G_7)$-exposed. From Lemma \ref{lem:upperbound} we have $r(G_7)\leq
4$. Thus $r(G_7)=4$.
\end{example}

\begin{table}[!ht]
  \caption{The entries of $f_0$ that are explicit in the proof of
    Lemma \ref{lem:upperbound}, the permutations in $G_7$ by which
    they are $3$-exposed, and the relevant $A_{i\mapsto f_0(i)}$
    sets.}
  \label{tab:G7f0Exp}
  \begin{center}
    \begin{tabular}{c|c|c}
      $f_0$ & $3$-exposed by & $A_{i\mapsto f_0(i)}$ \\ \hline $1 \mapsto 5$
      & $g^0$ & $A_{1\mapsto 5}=[1]=\mathset{1}$ \\
      $3 \mapsto 6$ & $g^5,g^6$ & $A_{3\mapsto 6}=[2,3]=\mathset{2,3}$\\
      $6 \mapsto 7$ & $g^2,g^3,g^4$ & $A_{6\mapsto 7}=[4,6]=\mathset{4,5,6}$ \\
      $5 \mapsto 1$ & $g^0,g^1,g^2$ & $A_{5\mapsto 1}=[6,8]\bmodp 7=\mathset{6,7,1}$ \\
    \end{tabular}
  \end{center}
\end{table}

The upper bound of Lemma \ref{lem:upperbound} and the upper bound of
Lemma \ref{lem:lowerbound} do not match exactly. The gap between the
two is eliminated in the following theorem, by improving the upper
bound, thus giving the exact covering radius of $G_n$.

\begin{theorem}
  \label{th:rgn}
  For all $n\in\N$,
  \[r(G_n) = n-\floorenv{\frac{\sqrt{4n+1}+1}{2}}.\]
\end{theorem}
\begin{IEEEproof}
  For $n=1,2$ we already know that $r(G_n)=0$, agreeing with the
  claimed expression. Therefore we consider $n\geq 3$. By Lemma
  \ref{lem:upperbound} and Lemma \ref{lem:lowerbound} we have
  \[ n-\floorenv{\frac{\sqrt{4n+1}+1}{2}} \leq r(G_n) \leq
  n- \ceilenv{\frac{\sqrt{4n+1}-1}{2}}.\] Using straightforward
  analysis, one can see that the lower and upper bounds agree, except
  when $n=t(t+1)$, $t\in\N$, where there is a gap of $1$ between the
  bounds. To prove the claim we shall strengthen the upper bound to
  match the lower bound.

  For the remainder of the proof we focus on the case of $n=t(t+1)$,
  $t\in\N$. In this case, there is no need for the floor or ceiling
  operations, and we would like to prove that
  \[ r(G_n)=n-\frac{\sqrt{4n+1}+1}{2} = t^2-1.\]
  Denote $\rt\eqdef t^2-1$, and assume to the contrary that there
  exists $f\in S_n$ that is $(\rt,G_n)$-exposed. Then,
  \[ n \overset{\text{(a)}}{=} \abs{\bigcup_{j\in[n]} A_{f^{-1}(j)\mapsto j}}
    \leq \sum_{j\in[n]} \abs{A_{f^{-1}(j)\mapsto j}}
    \overset{\text{(b)}}{=} (n-\rt-1)(n-\rt)=t(t+1)=n,
  \]
  where (a) follows from Lemma \ref{lem:entrycover}, and (b) is taken
  from \eqref{eq:union}. It follows that the sets $A_{f^{-1}(j)\mapsto
    j}$, $j\in[n]$, are all disjoint, and they form a partition of
  $[n]$.

  Define a \emph{$B$-set} to be any set of the form
  $A_{f^{-1}(j_B)\mapsto j_B}$, with $j_B\in B$, and a \emph{$T$-set}
  to be any set $A_{f^{-1}(j_T)\mapsto j_T}$, with $j_T\in T$. Since
  $\rt\geq \frac{n}{2}-1$, we have $B\cap T=\emptyset$, and thus no
  $B$-set is also a $T$-set. As noted above, the $B$-sets and $T$-sets
  partition $[n]$, and therefore there exists some $T$-set immediately
  to the left (cyclically) of a $B$-set. More precisely, there exist
  $j_B\in B$ and $j_T\in T$ such that
  \begin{align*}
    A_{f^{-1}(j_T)\mapsto j_T}&=[k,k+\ell_T] \bmodp n, \\
    A_{f^{-1}(j_B)\mapsto j_B}&=[k+\ell_T+1,k+\ell_T+\ell_B] \bmodp n,
  \end{align*}
  for some $k,\ell_B,\ell_T\in[n]$. But by Lemma \ref{lem:exposedset},
  \begin{align*}
    A_{f^{-1}(j_T)\to j_T} &= [f^{-1}(j_T)-j_T+\rt+2,f^{-1}(j_T)] \bmodp n,\\
    A_{f^{-1}(j_B)\to j_B} &= [f^{-1}(j_B)+1,f^{-1}(j_B)+n-\rt-j_B] \bmodp n,
  \end{align*}
  implying $f^{-1}(j_B)=f^{-1}(j_T)$, and therefore $j_B=j_T$, but then $B\cap
  T\neq\emptyset$, a contradiction.
\end{IEEEproof}


\section{Codes Constructed from the Cyclic Group}
\label{sec:code}

Using $G_n$ as a covering code, now that its covering radius has been
determined, has severe limitations. Most notably, there is just one
code of each length, and no flexibility in code parameters. We
overcome this by providing a more general code construction which uses
$G_n$ as an internal building block. This construction is a
generalization of the covering-code construction of
\cite{FarSchBru16a}. It enables us to construct a covering code
$C_n\subseteq S_n$ , using existing covering codes $C_m\subseteq S_m$,
$m\leq n$.


\subsection{Code Construction and Parameters}

Before describing the construction we first define permutation
projections.

\begin{definition}
  Let $I=\mathset{i_1,i_2,\dots,i_m}\subseteq [n]$ be a subset of
  indices, $i_1<i_2<\dots<i_m$. For a permutation $f\in S_n$ we define
  $f|_I$ to be the permutation in $S_m$ that preserves the relative
  order of the sequence $f(i_1),f(i_2),\dots,f(i_m)$, i.e., $g=f|_I$
  if for all $j,j'\in[m]$, we have $g(j)<g(j')$ if and only if
  $f(i_j)<f(i_{j'})$. We also define
  \[ f|^I \eqdef \parenv{f^{-1}|_I}^{-1}.\]
\end{definition}

Intuitively, from the definition above, to compute $f|_I$ we take its
one-line notation, keep only the \emph{coordinates} of $f$ from $I$,
and then rename them to the elements of $[m]$ while keeping the
relative order. In contrast, to compute $f|^I$, we keep only the
one-line notation \emph{values} of $f$ that are from $I$, and rename
those to $[m]$ while keeping the relative order.

\begin{example}
  Let $n=6$, $f=[6,1,3,5,2,4]\in S_6$, and $I=\mathset{3,5,6}$. Then
  \[f|_I=[2,1,3],\]
  since we keep entries $3$, $5$, and $6$ of $f$, giving us $[3,2,4]$,
  which we then rename to $[2,1,3]$. Similarly, we have
  \[f|^I=[3,1,2],\]
  since we keep the values $3$, $5$, and $6$ of $f$, giving us
  $[6,3,5]$, which we then rename to $[3,1,2]$.
\end{example}

To simplify notation, it will become convenient to define a projection
using the empty set. Thus, for $I=\emptyset$ and $f\in S_n$ we define
$f|_I=f|^I\eqdef []$, where $[]$ denotes the unique permutation over
zero elements.

We now present the code construction.

\begin{construction}
  \label{con:code}
  Let $m,n\in\N$, $m\leq n$. We define the indices sets
  \[ I_i\eqdef [im+1,(i+1)m]\cap [n],\]
  for all $i\in [0,\floorenv{\frac{n}{m}}]$. We construct the code
  $C_n\subseteq S_n$ defined by
  \[ C_n\eqdef \mathset{ f\in S_n ~:~ f|^{I_i}\in C_{\abs{I_i}}, i\in\sparenv{0,\floorenv{\frac{n}{m}}}},\]
  where $C_{\abs{I_i}}\subseteq S_{\abs{I_i}}$ are covering codes,
  called \emph{the building-block codes}.
\end{construction}

We note that in the above construction, all the indices sets are of
size $m$, except for the last one which is of size $n\bmod m$. Thus,
when $m|n$ the last indices set is empty, and
$C_0\eqdef\mathset{[]}=\mathset{\id}\subseteq S_0$ is degenerate,
containing only the unique empty (identity) permutation. We define
$r(C_0)\eqdef 0$. We also mention that a more general construction is
possible, in which the indices sets form an arbitrary partition of
$[n]$.

The code construction of \cite{FarSchBru16a} is a special case of
Construction \ref{con:code}, in which $C_m\eqdef\mathset{\id}\subseteq
S_m$, and $C_{n\bmod m}\eqdef\mathset{\id}\subseteq S_{n\bmod m}$.

\begin{lemma}
  \label{lem:codeparm}
  The code $C_n$ from Construction \ref{con:code} is an $(n,M,r)$ code,
  where
  \[ M=\frac{n!}{(m!)^{\floorenv{n/m}}(n\bmod m)!}\abs{C_m}^{\floorenv{n/m}}\abs{C_{n\bmod m}},\]
  and
  \[ r=\max\mathset{r(C_m),r(C_{n\bmod m})}.\]
\end{lemma}
\begin{IEEEproof}
  The cardinality of the code, $M$, is easily obtainable by noting that we
  first need to partition the $n$ coordinates into $\floorenv{\frac{n}{m}}$ sets
  of size $m$, and one set of size $n\bmod m$. There are
  \[ \binom{n}{m,m,\dots,m,n \bmod m}=\frac{n!}{(m!)^{\floorenv{n/m}}(n\bmod m)!}\]
  ways of doing so. We then assign values to each set from the
  corresponding set $I_i$. The number of ways to do so is exactly
  $\abs{C_m}^{\floorenv{n/m}}\abs{C_{n\bmod m}}$.

  The covering radius is also straightforward. Given a permutation
  $f\in S_n$, assume the values of $I_i$ are found in positions given
  by $J_i\subseteq [n]$. By the properties of the code
  $C_{\abs{I_i}}$, there exists a codeword $g\in C_n$, such that the
  restrictions of $f$ and $g$ to positions $J_i$ are at most
  $r(C_{\abs{I_i}})$ distance apart. Since we can make this hold for
  all $i\in[0,\floorenv{\frac{n}{m}}]$ simultaneously, we have
  \[ r\leq \max\mathset{r(C_m),r(C_{n\bmod m})}.\]
  This is met with equality, since we can easily find a permutation
  $f\in S_n$ within this distance from $C_n$: take $f'\in S_m$ such
  that $d(f',C_m)=r(C_m)$. Construct $f\in S_n$ such that
  $f|^{I_0}=f'$ and then $d(f,C_n)\geq r(C_m)$. If necessary, repeat
  analogously for $C_{n\bmod m}$ to obtain a permutation $f\in S_n$
  such that $d(f,C_n)\geq r(C_{n\bmod m})$.
\end{IEEEproof}

Next, we take a closer look at this code construction using $G_n$ as
the building block code.

\begin{corollary}
  \label{cor:gnparm}
  Let $m,n\in\N$, $m\leq n$. Then the code $C_n$ from Construction
  \ref{con:code}, with building-block codes $C_m=G_m$ and $C_{n\bmod
    m}=G_{n\bmod m}$, is an $(n,M,r)$ code, where
  \[ M=\begin{cases}
  \frac{n!}{((m-1)!)^{\frac{n}{m}}} & n\equiv 0 \pmod{m},\\
  \frac{n!}{((m-1)!)^{\floorenv{\frac{n}{m}}}((n\bmod m)-1)!} & n\not\equiv 0 \pmod{m},\\
  \end{cases}\]
  and
  \[ r=m-\floorenv{\frac{\sqrt{4m+1}+1}{2}}.\]
  Here we use the convention that $G_0=\mathset{[]}$.
\end{corollary}
\begin{IEEEproof}
  The proof follows from substituting the parameters of the cyclic
  group into Lemma \ref{lem:codeparm}, and noting that $r(G_m)$ is
  monotone non-decreasing in $m$.
\end{IEEEproof}

\begin{lemma}
  Let $n,m\in\N$, $m\leq n$. Then the code $C_n$ of Construction
  \ref{con:code} with $C_m=G_m$ and $C_{n\bmod
    m}=G_{n\bmod m}$, has the following rate,
  \begin{equation}
    \label{eq:rrho}
    R = -\rho\floorenv{\frac{1}{\rho}}\log_2 \rho-\parenv{1-\rho\floorenv{\frac{1}{\rho}}}\log_2\parenv{1-\rho\floorenv{\frac{1}{\rho}}}+o(1),
  \end{equation}
  where $\rho\eqdef \rho(C_n)$ is the normalized covering radius of
  $C_n$, $R\eqdef R(C_n)$ is the rate of $C_n$, and $o(1)$ denotes a
  function that tends to $0$ as $n$ tends to infinity.
\end{lemma}
\begin{IEEEproof}
  From Corollary \ref{cor:gnparm}
  \[ \rho=\frac{r(C_n)}{n-1}=\frac{m - \floorenv{\frac{\sqrt{4m+1}+1}{2}}}{n-1}=\frac{m}{n}-o(1). \]
  Therefore, $m = n\rho + o(n)$. Notice that $n\bmod m =
  n-m\floorenv{\frac{n}{m}}$, hence, by rewriting $\abs{C_n}$ from
  Corollary \ref{cor:gnparm} we get
  \begin{align*}
    \abs{C_n} = 2^{Rn} &= \frac{n!}{(m!)^{\floorenv{\frac{n}{m}}}(n\bmod m )!} m^{\floorenv{\frac{n}{m}}}(n\bmod m) \\
    &= \frac{n!}{((n\rho+o(n))!)^{\floorenv{\frac{n}{n\rho+o(n)}}}\parenv{n-(n\rho+o(n))\floorenv{\frac{n}{n\rho+o(n)}}}!} \\
    &\quad\  \cdot (n\rho+o(n))^{\floorenv{\frac{n}{n\rho+o(n)}}}\parenv{n-(n\rho+o(n))\floorenv{\frac{n}{n\rho+o(n)}}}.
\end{align*}
  It is now a matter of using Stirling's approximation (e.g.,
  \cite{GraKnuPat94}),
  \[ n! = \parenv{ \frac{n}{e} }^n 2^{o(n)},\]
  and standard analysis techniques, to arrive at the desired form.
\end{IEEEproof}

We observe that \eqref{eq:rrho} is the same as the rate obtained by
the construction of \cite{FarSchBru16a}, which uses only
$C_m=\mathset{\id}$. However, the rate is a rather crude measure. Upon
closer inspection, we shall now show the code parameters of Corollary
\ref{cor:gnparm} are superior to those of \cite{FarSchBru16a}.

To avoid clutter, let us consider the case of $n=tm$, where
$t,m\in\N$. We use Construction \ref{con:code} with $C_m=G_m$ to
obtain a code we denote as $\ccyc_n$. This code has cardinality
given by Corollary \ref{cor:gnparm},
\[ \Mcyc_n = \frac{(mt)!}{((m-1)!)^t}.\]
Its covering radius is
\[ r\eqdef r(\ccyc_n) = m-\floorenv{\frac{\sqrt{4m+1}+1}{2}}.\]
For a fair comparison with the code of \cite{FarSchBru16a}, we
construct one with the same length $n$, and same covering radius
$r$. Such a code is a special case of Construction \ref{con:code}
using the building-block codes $C_{r+1}=\mathset{\id}$ and $C_{n\bmod
  (r+1)}=\mathset{\id}$. We call the resulting code $\cid_n$, and its
cardinality (see also \cite{FarSchBru16a}) is given by
\[ \Mid_n=\frac{(mt)!}{((r+1)!)^{\floorenv{\frac{n}{r+1}}}(n\bmod(r+1))!}.\]

For the comparison, we first observe that
\begin{equation}
  \label{eq:sqrt}
  r\leq m-\sqrt{m}+1.
\end{equation}
We also recall Stirling's approximation in more detail,
\begin{equation}
  \label{eq:stir}
  \sqrt{2\pi n}\parenv{\frac{n}{e}}^n\leq n!\leq \sqrt{2\pi n}\cdot e^{\frac{1}{12n}}\parenv{\frac{n}{e}}^n.
\end{equation}
We now have
\[  \Mcyc_n = \frac{(tm)!\cdot m^t}{(m!)^t}
  \overset{\text{(a)}}{\leq} \frac{\sqrt{2\pi tm} \parenv{\frac{tm}{e}}^{tm}e^{\frac{1}{12tm}}\cdot m^t} {\parenv{\frac{m}{e}}^{tm}(2\pi m)^{\frac{t}{2}}}
  \overset{\text{(b)}}{\leq} 2\sqrt{t} \cdot m^t t^{tm},
\]
where (a) is obtained by using \eqref{eq:stir}, and (b) is by
rearrangement and noting that $e^{\frac{1}{12tm}}\leq 2$.

To bound $\Mid$ we write
\[ n=tm=q(r+1)+s,\]
where $q,s\in\Z$, $s\in [0,r]$. We then have
\begin{align*}
  \Mid_n &= \frac{(tm)!}{((r+1)!)^q \cdot s!} \\
  &\overset{\text{(a)}}{\geq} \frac{\sqrt{2\pi tm}\parenv{\frac{tm}{e}}^{tm}} {(2\pi(r+1))^{\frac{q}{2}} e^{\frac{q}{12(r+1)}} \parenv{\frac{r+1}{e}}^{(r+1)q} \cdot (2\pi s)^{\frac{1}{2}} e^{\frac{q}{12(r+1)}+\frac{1}{12s}}\parenv{\frac{s}{e}}^s}\\
  &\overset{\text{(b)}}{\geq} \frac{t^{tm}}
  {\parenv{\frac{r+1}{m}}^{(r+1)q}\parenv{\frac{s}{m}}^s 2^{2t+1}(2\pi tm)^t} \\
  &\overset{\text{(c)}}{\geq} \frac{t^{tm}}
  {\parenv{\frac{r+1}{m}}^{tm} 2^{2t+1}(2\pi tm)^t} \\
  &\overset{\text{(d)}}{\geq} \frac{t^{tm}}
  {\parenv{1-\frac{1}{\sqrt{m}}+\frac{2}{m}}^{tm} 2^{2t+1}(2\pi tm)^t}\\
  &\overset{\text{(e)}}{\geq} \frac{t^{tm}}
  {\parenv{e^{-\sqrt{m}+2}}^{t} 2^{2t+1}(2\pi tm)^t},
\end{align*}
where (a) is due to \eqref{eq:stir}, (b) is by rearrangement and
noting that $q\leq 2t$, (c) is due to $s\leq m$, (d) is due to
\eqref{eq:sqrt}, and (e) is due to $1+x\leq e^x$. It now follows that
\[ \frac{\Mcyc}{\Mid} \leq 2^{2t+2}\sqrt{t}(2\pi t)^t \parenv{m^2 e^{-\sqrt{m}+2}}^t .\]
Thus, for any fixed $t\in\N$, and $m$ tending to infinity, the codes
$\ccyc_n$ are sub-exponentially better than $\cid_n$ of
\cite{FarSchBru16a} in terms of size.

As a final note, we mention the fact that we may improve the
parameters of Corollary \ref{cor:gnparm} by picking $C_m=G_m$, but
$C_{n\bmod m}=\mathset{\id}$, whenever $(n\bmod m)-1\leq r(G_m)$, as
this would decrease the resulting code size while maintaining its
covering radius.


\subsection{Covering-Codeword Algorithm}

A common task associated with covering codes is, given a covering code
$C\subseteq S_n$ and a permutation $f\in S_n$, to find a codeword
$g\in C$ such that $d(f,g)\leq r(C)$, i.e., find a codeword covering
$f$. The code $G_n$ is small, and a trivial algorithm measuring the
distance between the given $f$ and each of the $n$ codewords of $G_n$
(returning an $r(G_n)$-covering codeword) runs in $O(n^2)$
time. However, this might be improved upon, and we now describe a more
efficient algorithm.

\begin{algorithm}[ht]
\caption{Finding a covering codeword $g\in G_n$}
\label{alg:CycFind}
\begin{algorithmic}
\STATE \textbf{Input:} any permutation $f \in S_n$
\STATE \textbf{Output:} a codeword $g\in G_n$ with $d(f,g)\leq r(G_n)$
\STATE \textbf{Initialization:} $V$ is an array of size $n$, $V[i]\leftarrow 0,\ \forall i\in [n],\ a\leftarrow \floorenv{\frac{\sqrt{4n+1}-1}{2}}$
\FOR{ $i=1$ to $n$}
\IF{ $f(i) \leq a$}
\FOR{ $j=i+1$ to $i+a-(f(i)-1)$}
\STATE $V \sparenv{ j\bmodp n } \leftarrow 1$
\ENDFOR
\ELSIF{ $f(i)\geq n-a+1$}
\FOR{ $j=i-(a-(n-f(i)))+1$ to $i$}
\STATE $V \sparenv{ j\bmodp n } \leftarrow 1$
\ENDFOR
\ENDIF
\ENDFOR
\FOR{ $i=1$ to $n$ }
\IF{ $V[i]=0$ }
\RETURN{$[n-i+2,\dots,n,1,\dots,n-i+1]\in G_n$}
\ENDIF
\ENDFOR
\end{algorithmic}
\end{algorithm}

\begin{lemma}
\label{lem:CycFind}
Let $n\in \N$ and $f\in S_n$. Algorithm \ref{alg:CycFind} returns a
codeword $g\in G_n$ such that $d(f,g)\leq r(G_n)$.
\end{lemma}
\begin{IEEEproof}
  Let $\rt\eqdef r(G_n)$, which means $a=n-\rt-1$. The
  inner loops on $j$ assign $1$ to the entries of $V$ corresponding to
  the elements of $A_{i\mapsto f(i)}$ (see proof of Lemma
  \ref{lem:lowerbound}). Hence, at the end of the first for loop on
  $i$,
  \[ V[i]=0 \iff i \notin \bigcup_{i\in[n]}A_{i\mapsto f(i)}. \]
  The second for loop on $i$ finds $i\in [n]$ such that $V[i]=0$. From
  Theorem \ref{th:rgn}, such $i$ must exist. We conclude that the
  codeword $g\in G_n$, such that $g(i)=1$, $\rt$-covers $f$, and we
  return it.
\end{IEEEproof}

Algorithm \ref{alg:CycFind} is more efficient than the trivial
brute-force algorithm. We note that $a=O(\sqrt{n})$, and therefore,
each of the inner loops is entered $O(\sqrt{n})$ times, performing
$O(\sqrt{n})$ iterations each time.  Thus, in total, the algorithm
runs in $O(n)$ time.

Having this algorithm for the building-block code $G_n$, we may extend
it in a natural way to the code studied in Corollary \ref{cor:gnparm}
to also run in $O(n)$ time. We omit the tedious details.


\section{Relabeling the Cyclic Group}
\label{sec:relabel}

Following the definition of the \emph{natural} transitive cyclic
group,
\[G_n\eqdef\spn{(1,2,\dots,n)}\subseteq S_n,\]
as given in Definition \ref{def:natgn}, it is tempting to ask what
happens when we take a non-natural transitive cyclic group. Thus,
we are interested in the groups of the form
\[ G_n^h\eqdef h G_n h^{-1} \eqdef \spn{h(1,2,\dots,n)h^{-1}}=\spn{(h(1),h(2),\dots,h(n))}\subseteq S_n,\]
for some $h\in S_n$. A similar, more general question, was asked in
\cite{TamSch12}, where an error-correcting code $C\subseteq S_n$ was
\emph{relabeled} by conjugation,
\[C^{h}\eqdef h Ch^{-1}\eqdef \mathset{h g h^{-1} ~:~ g\in C},\]
$h\in S_n$, and its minimum distance was studied as a function of
$C$ and $h$. It was shown there that the minimum distance could
drastically change due to relabeling, moving from the minimum possible
$1$, to the maximum possible $n-1$, for some codes. Additionally,
every error-correcting code could be relabeled so that its minimum
distance is reduced to either $1$ or $2$. In this section we study the
covering radius of relabelings of $G_n$.

\begin{definition}
  Let $C\subseteq S_n$ be a covering code. We denote by $\lmin(C)$
  (respectively, $\lmax(C)$) the minimal (respectively, maximal) achievable
  covering radius among all relabelings of $C$, i.e.,
  \begin{align*}
    \lmin(C) &\eqdef \min_{h\in S_n} r(C^h), \\
    \lmax(C) &\eqdef \max_{h\in S_n} r(C^h).
  \end{align*}
\end{definition}

We first consider $\lmax(G_n)$. Again, the cases of $n=1,2$ are degenerate,
and we therefore only consider $n\geq 3$.

\begin{theorem}
  \label{th:lmaxgn}
  For all $n\in\N$, $n\geq 3$,
  \[ \lmax(G_n)=n-\ceilenv{\frac{\sqrt{4n+1}-1}{2}}.\]
\end{theorem}
\begin{IEEEproof}
  Let $h\in S_n$ be any permutation.  We begin by noting that since
  $G_n$ is a transitive group, so is $G_n^h$. Thus, Lemma
  \ref{lem:entrycover} and Lemma \ref{lem:exposedset} apply. Now Lemma
  \ref{lem:upperbound} also holds for $G_n^h$ since it only relies on
  the two above-mentioned lemmas. Thus,
  \[ \lmax(G_n)\leq n-\ceilenv{\frac{\sqrt{4n+1}-1}{2}}.\]
  Additionally, whenever $n\neq t(t+1)$, $t\in\N$, we have by
  Theorem \ref{th:rgn}
  \[ \lmax(G_n)\geq r(G_n)=n-\floorenv{\frac{\sqrt{4n+1}+1}{2}}=n-\ceilenv{\frac{\sqrt{4n+1}-1}{2}}.\]

  Let us define
  \begin{align*}
    a&\eqdef\ceilenv{\frac{\sqrt{4n+1}-1}{2}}, & \rt&\eqdef  n-a-1.
  \end{align*}
  To complete this proof, we must show that for values of $n$ such
  that $n=t(t+1)$, $t\in \N$, $t \geq 2$, there exists $h\in S_n$ such
  that $r(G_n^h)=n-a$. Notice that in this case,
  $\frac{\sqrt{4n+1}-1}{2}$ is an integer, which yields $n = a(a+1)$.

  We contend that the permutation $h\eqdef (1,2)\in S_n$ will suffice,
  proving it by constructing a permutation $f_0\in S_n$ such that
  $f_0$ is $(\rt,G_n^h)$-exposed, giving us
  \[ r(G_n^h)\geq d(f_0,G_n^h)\geq \rt+1 = n-a.\]
  
  We construct a permutation $f_0\in S_n$ as follows:
  \begin{align}
    \label{eq:f0l}
    f_0(i) &\eqdef
    \begin{cases}
      1 & i=1,\\
      n & i=2,\\
      n-a+1 & i=3,\\
      a-k & i=\binom{k+1}{2}+a+2, k\in [0,a-2], \\
      n-a+1+\ell & i=n-a+2-\binom{\ell+1}{2}, \ell\in [a-2], \\
      \text{arbitrary} &\text{otherwise},
    \end{cases}
  \end{align} 
  for all $i\in[n]$, and where arbitrary entries are set in a way that
  completes $f_0$ to a permutation.

  We first note that $f_0$ is well defined. The domain intervals in
  the definition are disjoint since $a\geq 2$, $n=a(a+1)=2\binom{a+1}{2}$,
  and
  \[ \binom{a-1}{2}+a+2 < 2\binom{a+1}{2}-a+2-\binom{a-1}{2}.\]
  As for the range intervals, the fourth and fifth cases in
  \eqref{eq:f0l} are $[2,a]$ and $[n-a+2,n-1]$ respectively, and are
  clearly disjoint, and disjoint from the first three cases. These two sets
  will be of further interest, so we define
  \begin{align*}
    \Bt &\eqdef B\setminus\mathset{1} = [2,a],\\
    \Tt &\eqdef T\setminus\mathset{n-a+1,n} = [n-a+2,n-1].
  \end{align*}
  Thus, $\Bt\cap\Tt=\emptyset$.

  With $g\eqdef (1,2,\dots,n)\in S_n$, and $G_n\eqdef\spn{g}$, we write the
  elements of $G^h_n$ explicitly,
  \begin{align*}
    h_0&\eqdef hg^0h^{-1} = [1,2,\dots,n],\\
    h_1&\eqdef hg^1h^{-1} = [3,1,4,5,\dots,n,2],\\
    h_2&\eqdef hg^2h^{-1} = [4,3,5,6,\dots,n,2,1],\\
    h_i&\eqdef hg^{i}h^{-1} = [i+2,i+1,i+3,i+4,\dots,n,2,1,3,4,\dots,i], i\in[3,n-3],\\
    h_{n-2}&\eqdef hg^{n-2}h^{-1} = [n,n-1,2,1,3,4,\dots,n-2],\\
    h_{n-1}&\eqdef hg^{n-1}h^{-1} = [2,n,1,3,4,\dots,n-1].
  \end{align*}
  To prove that $f_0$ is $(\rt,G^h_n)$-exposed we shall use Lemma
  \ref{lem:entrycover}. 

  The mapping $1\mapsto f_0(1)=1$ is $\rt$-exposed by
  $\mathset{h_{n-a-1},h_{n-a},\dots,h_{n-2}}$, hence,
  \[ A_{1\mapsto 1}^{G^h_n}= [4,a+3].\]
  The mapping $2\mapsto f_0(2)=n$ is $\rt$-exposed by
  $\mathset{h_0,h_1,\dots,h_{a-1}}$, hence
  \[A_{2\mapsto n}^{G^h_n}=[n-a+3,n+2]\bmodp n=\mathset{n-a+3,n-a+4,\dots,n,1,2}.\]
  The mapping $3\mapsto f_0(3)=n-a+1$ is $\rt$-exposed solely by $h_{n-1}$,
  thus
  \[ A_{2\mapsto n-a+1}^{G^h_n}=\mathset{3}.\]
  Now consider a mapping $i_B\mapsto f_0(i_B)=j_B$, with $j_B\in \Bt$,
  and we get
  \[ A^{G_n^h}_{i_B \mapsto j_B}=[i_B+2,i_B+2+a-j_B],\]
  and in total,
  \[ \bigcup_{j_B\in \Bt}A^{G^h_n}_{f^{-1}_0(j_B)\mapsto j_B}=\sparenv{a+4,\binom{a+1}{2}+3}=\sparenv{a+4,\frac{n}{2}+3}. \]
  Similarly, for $i_T\mapsto f_0(i_T)=j_T$ such that $j_T \in \Tt$ we get
  \[ A^{G_n^h}_{i_T \mapsto j_T}=[i_T+n-j_T-a+2,i_T+1],\]
  and in total,
  \[ \bigcup_{j_T\in \Tt}A^{G^h_n}_{f^{-1}_0(j_T)\mapsto j_T}=\sparenv{n-\binom{a+1}{2}+4,n-a+2}=\sparenv{\frac{n}{2}+4,n-a+2}. \]
  In conclusion, taking the union of all the above we obtain
  \[ \bigcup_{j\in [n]}A^{G_n^h}_{f_0^{-1}(j)\mapsto j} =[n], \]
  and by Lemma \ref{lem:entrycover} we have that $f_0$ is
  $(\rt,G^h_n)$-exposed.
\end{IEEEproof}

We now move on to studying $\lmin$. Unlike $\lmax$, we provide only a
weak lower bound on $\lmin$, which depends only on the size of the
code. We recall the definition of a ball of radius $r$ and centered at
$g\in S_n$,
\[ \cB_{n,r}(g) \eqdef \mathset{ f\in S_n ~:~ d(f,g)\leq r}.\]
Since the $\ell_\infty$-metric is right invariant, the size of a ball
does not depend on the choice of center, and thus we denote its size
as $\abs{\cB_{n,r}}$.

\begin{lemma}
  \label{lem:trivlmin}
  Let $C\subseteq S_n$ be a code. If $\rt\in\N$ is such that
  \begin{equation}
    \label{eq:notcover}
    \abs{C}\cdot\abs{\cB_{n,\rt-1}} < \abs{S_n},
  \end{equation}
  then
  \[ \lmin(C) \geq \rt.\]
\end{lemma}
\begin{IEEEproof}
  The claim is quite trivial. Inequality \eqref{eq:notcover} simply
  states that $\abs{C}$ balls of radius $\rt-1$ cannot cover $S_n$,
  hence $r(C)\geq \rt$. For all $h\in S_n$ we have
  $\abs{C}=\abs{C^h}$, hence $r(C^h)\geq \rt$.
\end{IEEEproof}

Specializing Lemma \ref{lem:trivlmin} to $\abs{C}=n$, gives us the
following corollary, which applies to $G_n$ as well.

\begin{corollary}
  For all large enough $n\in\N$, $C\subseteq S_n$, $\abs{C}=n$,
  \[ \lmin(C)\geq n-\ceilenv{\sqrt{2n\ln n+2n}}.\]
\end{corollary}
\begin{IEEEproof}
The following upper bound on the size of a ball is given in
\cite{Klo08},
\[ \abs{\cB_{r,n}} \leq 
\begin{cases}
((2r+1)!)^{\frac{n-2r}{2r+1}}\prod_{i=r+1}^{2r}(i!)^{\frac{2}{i}} & 0\leq r \leq\frac{n-1}{2}, \\
(n!)^{\frac{2r+2-n}{n}}\prod_{i=r+1}^{n-1}(i!)^{\frac{2}{i}} & \frac{n-1}{2}\leq r \leq n-1,
\end{cases} \]
and whose proof is an immediate application of Bregman's upper bound
on the permanent. We contend that only the second case of this bound
is of relevance to us, as we will prove shortly. Thus, if we find
$\rt\geq \frac{n+1}{2}$ such that
\begin{equation}
  \label{eq:lminrt}
  \frac{\abs{C}\cdot\abs{\cB_{n,\rt-1}}}{\abs{S_n}}
  = \frac{\abs{\cB_{n,\rt-1}}}{(n-1)!}\leq \frac{1}{(n-1)!}(n!)^{\frac{2\rt-n}{n}}\prod_{i=\rt}^{n-1}(i!)^{\frac{2}{i}} < 1,
\end{equation}
then by Lemma \ref{lem:trivlmin} we will have $\lmin(C)\geq \rt$.

Let us therefore define the auxiliary function,
\[ F(n,\rt)\eqdef \frac{1}{(n-1)!}(n!)^{\frac{2\rt-n}{n}}\prod_{i=\rt}^{n-1}(i!)^{\frac{2}{i}}.\]
As a first step we show that for all $n\geq 11$,
\[ F\parenv{n,\ceilenv{\frac{n+1}{2}}} < 1.\]
Due to parity, we consider the cases of even $n$ and odd $n$
separately. We shall prove the former, and omit the proof for odd $n$
since it is similar. For the case of even $n$, we prove the claim for
$n=12$, and then show the function is monotonically decreasing in $n$.

For $n=12$ we have,
\[ F(12,7) \approx 0.9644 < 1.\]
Next, we consider
\begin{align*}
  \frac{F\parenv{n,\frac{n+2}{2}}}{F\parenv{n+2,\frac{n+4}{2}}} &= \frac{n\cdot(n!)^{\frac{2-n}{n}}\cdot\prod_{i=\frac{n+2}{2}}^{n-1}(i!)^{\frac{2}{i}}}{(n+2)\cdot((n+2)!)^{-\frac{n}{n+2}}\cdot\prod_{i=\frac{n+4}{2}}^{n+1}(i!)^{\frac{2}{i}}} \\
  &= \frac{n(n+1)\cdot \parenv{\parenv{\frac{n+2}{2}}!}^{\frac{4}{n+2}}}{((n+2)!)^{\frac{2}{n+2}}\cdot ((n+1)!)^{\frac{2}{n+1}}} \\
  &\geq \frac{e^2}{4}\cdot \frac{n}{(n+1)\cdot (\pi (n+2))^{\frac{1}{(n+1)(n+2)}}\cdot e^{\frac{1}{6}\parenv{\frac{1}{(n+1)^2}+\frac{1}{(n+2)^2}}}},
\end{align*}
where for the inequality we used \eqref{eq:stir} and trivial bounding
techniques. We now note that
$\exp(\frac{1}{6}\parenv{\frac{1}{(n+1)^2}+\frac{1}{(n+2)^2}})$ and
$(\pi (n+2))^{\frac{1}{(n+1)(n+2)}}$ are monotonically decreasing in
$n$, and $\frac{n}{n+1}$ is monotonically increasing. Hence,
\[ \frac{F\parenv{n,\frac{n+2}{2}}}{F\parenv{n+2,\frac{n+4}{2}}}
\geq \frac{F\parenv{12,7}}{F\parenv{14,8}}\approx 1.649 >1,\] and so
$F\parenv{n,\ceilenv{\frac{n+1}{2}}}$ is monotonically decreasing in
$n$ for even $n$. A similar proof holds for odd $n$.

Thus far we showed there exists $\rt\geq \frac{n+1}{2}$ that satisfies
\eqref{eq:lminrt} (in particular, $\rt=\ceilenv{(n+1)/2}$ does).  We
would now like to find such $\rt$ as large as possible. We observe the
following sequence of inequalities, where we take $n\geq 1$, and
$\frac{n+1}{2}\leq \rt\leq n-1$.
\begin{align}
  F(n,\rt) &\eqdef \frac{1}{(n-1)!}(n!)^{\frac{2\rt-n}{n}}\prod_{i=\rt}^{n-1}(i!)^{\frac{2}{i}}\nonumber\\
  &\overset{\text{(a)}}{\leq} n\cdot \parenv{\frac{n}{e}}^{2\tilde{r}-2n}\cdot \prod_{i=\tilde{r}}^{n-1}(2\pi i)^{\frac{1}{i}}e^{\frac{1}{6i}}\parenv{\frac{i}{e}}^2\nonumber\\
  &\overset{\text{(b)}}{\leq} n\cdot n^{2\tilde{r}-2n}\cdot (2\pi \tilde{r})^{\frac{n-\tilde{r}}{\tilde{r}}}e^{\frac{n-\tilde{r}}{6\tilde{r}}} \parenv{\frac{(n-1)!}{(\tilde{r}-1)!}}^2\nonumber\\
  &\overset{\text{(c)}}{\leq} n\cdot n^{2\tilde{r}-2n}\cdot \pi n e^{\frac{1}{6}}\cdot \parenv{\frac{(n-1)!}{(\tilde{r}-1)!}}^2\nonumber\\
  &\overset{\text{(d)}}{\leq} \pi e^{\frac{1}{6}} e^{\frac{1}{6(n-1)}} e^{2\tilde{r}-2n} n^2\cdot \frac{\parenv{\frac{n-1}{n}}^{2n}\frac{1}{n-1}}{\parenv{\frac{\tilde{r}-1}{n}}^{2\tilde{r}}\frac{1}{\tilde{r}-1}}\nonumber\\
  &\overset{\text{(e)}}{\leq} \pi e^{-\frac{109}{60}} e^{2\tilde{r}-2n} n^2\cdot \frac{1}{\parenv{\frac{\tilde{r}-1}{\tilde{r}}}^{2\tilde{r}}\parenv{\frac{\tilde{r}}{n}}^{2\tilde{r}}}\nonumber\\
  \label{eq:last}
  &\overset{\text{(f)}}{\leq} \pi e^{-\frac{109}{60}} \parenv{\frac{6}{5}}^{12}\cdot e^{2\tilde{r}-2n} n^2 \parenv{\frac{n}{\tilde{r}}}^{2\tilde{r}},
\end{align}
where (a) follows from \eqref{eq:stir}, (b) follows by noting that
$(2\pi i)^{\frac{1}{i}}$ and $e^{\frac{1}{6i}}$ are decreasing in $i$
and then replacing $i$ by $\rt$, (c) follows by noting that
$(2\pi\rt)^{\frac{n-\rt}{\rt}}$ and $e^{\frac{n-\rt}{6\rt}}$ are
decreasing in $\rt$ and replacing $\rt$ by $\frac{n}{2}$, (d) follows
again by use of \eqref{eq:stir}, (e) follows by noting that
$\exp(\frac{1}{6(n-1)})$ is decreasing in $n$ and substituting $n=11$,
that $((n-1)/n)^{2n}\leq e^{-2}$, and that $\frac{\rt-1}{n-1}<1$, and
finally, (f) follows by noting that $((\rt-1)/\rt)^{2\rt}$ is
increasing in $\rt$ and replacing $\rt$ (since $n\geq 11$ and
$\rt\geq\frac{n}{2}$) by $\rt=6$.

We note that taking $\tilde{r}=n-\sqrt{2n\ln{n}+2n}$, by \eqref{eq:last}
we get
\[ \lim_{n\to\infty}F(n,n-\sqrt{2n\ln{n}+2n}) \leq  \pi e^{-\frac{109}{60}} \parenv{\frac{6}{5}}^{12} \frac{1}{e^2} < 1.\]
It now follows that for large enough $n$,
\[F(n,n-\sqrt{2n\ln{n}+2n}) < 1,\]
and then
\[ \lmin(C)\geq n-\ceilenv{\sqrt{2n\ln n+2n}},\]
as claimed.
\end{IEEEproof}


\section{Discussion}
\label{sec:disc}

In this paper we found the exact covering radius of the (natural)
transitive cyclic group, $G_n$, and used it to construct new covering
codes. These codes often exhibit better parameters than known
covering-code constructions, while still allowing a linear-time
covering-codeword algorithm.

The methods we described may be extended to larger groups, e.g., the
dihedral group, though at a cost of a growing gap between the lower
and upper bounds on the covering radius. Thus, in the case of the
(naturally labeled) dihedral group, $D_n\leq S_n$, defined by,
\[ D_n \eqdef \spn{ (1,2,\dots,n),\prod_{i=1}^{\floorenv{n/2}}(i,n-i)},\]
we can obtain
\[ n-\floorenv{\frac{\sqrt{4n+1}+1}{2}} \geq r(D_n)\geq \begin{cases}
  n-\ceilenv{\frac{\sqrt{288n+297}-3}{16}} & n\in [4,9],\\
  n-\ceilenv{\frac{\sqrt{288n+737}-1}{16}} & n\in [10,911],\\
  n-\ceilenv{\frac{\sqrt{18n-18}}{4}} & n\geq 912.
\end{cases}
\]
The tedious proof follows the same logic as that presented in Section
\ref{sec:cover}, and the interested reader may find it in \cite{Kar16}. We
believe a more elegant treatment is needed.

Another gap exhibited in this work is between $\lmin(G_n)$ and
$\lmax(G_n)$. First, we note an interesting contrast with the case of
error-correcting codes (as described in \cite{TamSch12}). When
relabeling error-correcting codes, the minimum distance of \emph{any}
code, including $G_n$, may be reduced to either $1$ or $2$. The
minimum distance of $G_n$ is $\ceilenv{n/2}$, and the best possible
minimum distance after relabeling is
$n-\ceilenv{\frac{\sqrt{4n-3}-1}{2}}$, which bears a striking
resemblance to $r(G_n)$.

In light of Section \ref{sec:cover} and Section \ref{sec:relabel}, it
appears that the covering radius of $G_n$ and its conjugate, has much
less variance. This is evident from the small gap between $\lmin(G_n)$
and $\lmax(G_n)$, not to mention the fact that $r(G_n)=\lmax(G_n)$ in
most cases. We ran brute-force computer search, checking all possible
relabelings of $G_n$, $n\in[3,10]$. For this range,
\[ \lmin(G_n)=r(G_n)=\lmax(G_n),\]
for all $n\in[3,10]\setminus\mathset{6}$, and
\[ \lmin(G_6)=r(G_6)=\lmax(G_6)-1,\]
where of the $6!$ labeling permutations, $264$ give covering radius
$3=r(G_n)$, and $456$ give covering radius $4=\lmax(G_n)$. The gap
between $r(G_6)$ and $\lmax(G_6)$ is a consequence of Theorem
\ref{th:lmaxgn}. It is now tempting to conjecture that for all
$n\in\N$, $\lmin(G_n)=r(G_n)$. We leave this conjecture, and the
determination of the covering radius of other groups, as open
questions for future work.

\bibliographystyle{IEEEtranS}
\bibliography{allbib}

\end{document}